\pdfoutput=1
\documentclass{JINST}

\title{Measurements of the position-dependent photo-detection
  sensitivity of the Hamamatsu R11410 and R8520 photomultiplier tubes}

\author{L.~Baudis, 
   S.~D'Amato\footnote{Corresponding author: sandro@physik.uzh.ch},
    G.~Kessler, A.~Kish, J.~Wulf\\
 Physik-Institut, Universit\"at Z\"urich, \\
Winterthurerstr. 190, CH-8057, Switzerland}

\abstract{The Hamamatsu R11410 and R8520 photomultiplier tubes (PMTs) are designed for applications in dark matter detectors using liquid xenon, featuring excellent sensitivity to VUV light and
  stable operability at cryogenic temperatures. 
  For eleven R11410 and seven R8520 PMTs, we measured the relative photo-detection sensitivity at 470\,nm  as a function of the position of the incident light on the photocathode. 
  Considering 80\% of the photocathode
  surfaces, the observed   non-uniformity values are in the ranges of (5--10)\% and (25--30)\% for the R11410 and R8520 models, respectively.
We found that the non-uniformity in the inner region of the photocathode surface  is dominated by light reflections on the internal components  of the photosensors and that 
  the border regions contribute dominantly to the observed non-uniformity.}


\usepackage{float}
\usepackage{caption}
\usepackage{subcaption}
\usepackage{amsmath}
\usepackage[normalem]{ulem}

\begin{document}
\section{Introduction}
\label{chap:intro}
Photomultiplier tubes (PMTs) are commonly used as photosensors in
particle physics experiments due to their high quantum efficiency (QE), fast timing
characteristics, low dark current and single-photon sensitivity. Dark matter search
experiments using liquid xenon as the scattering target employ arrays of PMTs, exploiting their stable
operation in cryogenic environments~\cite{markpaper}. The distribution  of the observed VUV scintillation light in the PMT arrays is used
to reconstruct the event vertex. 
Generally, PMTs may show a non-uniform response depending on the hit
position of the incident light on the photocathode. 
The event position reconstruction relies on the individual
PMT response, hence an accurate characterisation of the PMTs is required. 
This may include the dependency of the sensitivity on the location of
the photoelectron emission.

This article describes the Hamamatsu R11410 (a \mbox{3-inch} diameter, circular
PMT) and the R8520 (a square \mbox{1-inch} PMT)~\cite{hama, hamamatsu}.  They
are designed to feature low radioactivity levels and optimised for the operation in
direct dark matter detection experiments using liquid xenon. They show a high QE at the
scintillation wavelength of xenon, which lies in the vacuum ultraviolet (VUV) region
 at 175~nm.  The R11410 is employed in the PandaX~\cite{panda}
and in the XENON1T~\cite{XE1T} detectors. Furthermore it will also be installed in next-generation dark
matter experiments, such as LZ~\cite{LZ} and XENONnT.  The R8520 PMT was employed
in the XENON10~\cite{XE10}, XENON100~\cite{XE100,225_days} and PandaX detectors, 
as well as in XENON1T and LZ for the surrounding xenon region.

The high relevance of characterising these PMTs already lead  to a vast
variety of studies and measurements. The general performance of the R11410 is
covered in~\cite{K.Lung}, its performance in cryogenic xenon environments
in~\cite{markpaper}, and its intrinsic radioactivity was evaluated
in~\cite{Akerib}. Its quantum efficiency was measured by~\cite{alexey}, and
the wavelength-dependent double photoelectron emission by~\cite{DPE}. 

This work aims to improve upon existing measurements of the position dependent
sensitivity of these PMTs. While \cite{hamamatsu, K.Lung} 
measured the photocathode current, we used a different approach and in addition, 
we increased the spatial resolution of the measurements. A position-dependent PMT response 
was observed, as a combined effect from the QE, the collection efficiency (CE) and the inner light reflections on the internal components of the tubes. The measured \emph{relative photo-detection sensitivity}
values were normalised to the maximum measured value for each PMT.
In section~\ref{chap:description}, we give a brief description of the PMTs and
introduce the experimental setup used in this work. The
analysis and the results of the performed measurements are presented in
section~\ref{chap:results}. Section~\ref{chap:conclusion} summarises the
main findings and provides an outlook.


\section{Experimental description}
\label{chap:description}

In this section we first give a short description of the tested PMTs. It is followed by a description of the experimental setup operated at the University of Z\"urich to
quantify the cathode uniformity of two models of Hamamatsu PMTs.

\subsection{Description of the tested PMTs} 

\begin{figure}[tbp]
  \centering
  \includegraphics[height=8cm]{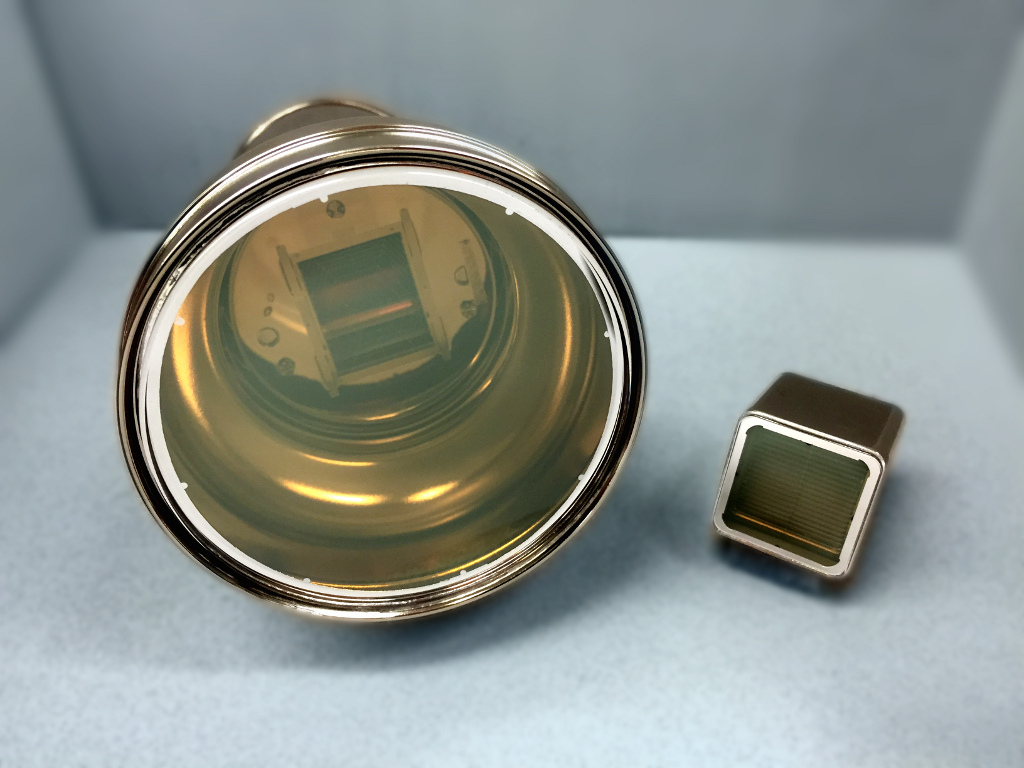}
  \caption{\small{(Left): Front view of a Hamamatsu R11410 PMT with visible inner
    structure (focusing dynode, first dynode and metal screws). (Right): View of a Hamamatsu R8520 PMT with visible first dynode stage.}}
  \label{pmtfoto}
\end{figure}

The Hamamatsu R11410 is a photomultiplier tube with a circular  3-inch diameter window \cite{hama, hamamatsu}, as shown in Figure \ref{pmtfoto}. It is specifically designed for low temperature operation (down to $-100^\circ$C) such as in liquid xenon. The wavelength-dependent quantum efficiency (QE) of its bialkali photocathode has a local maximum of $\sim$30 \% at 175 nm \cite{K.Lung}, the scintillation wavelength of xenon. The PMT window, made of synthetic silica, is transparent at this wavelength. The photoelectron collection efficiency (CE) on the first dynode is about 90 \% \cite{K.Lung, Akerib}.  
The bias voltage is around 1500\,V, with the maximum being 
1750\,V~\cite{markpaper}.  At 1500\,V the \mbox{12-dynode} PMT has an average gain of 5$\times 10^6$. It consists of a cobalt free Kovar metal body of 11.4\,cm length and a maximal (minimal) diameter of 7.8\,cm
(5.3\,cm). The photocathode has a minimum effective diameter of 6.4\,cm.

The Hamamatsu R8520 is a photomultiplier with a square 1-inch window \cite{hama, hamamatsu} (see Figure~\ref{pmtfoto}). The bialkali photocathode has a spectral response from 160 to 650\,nm wavelength with a quantum efficiency of 30 \% \cite{arXiv1207} at 175\,nm.  The \mbox{10-dynode} PMT with a typical gain of 2$\times10^6$~\cite{XE100} and dimensions of $25.7\times25.7\times28.2$ mm$^{3}$ has an active area of the
photocathode of $20.5\times20.5$ mm$^{2}$. The voltage applied between anode and cathode is around 800\,V, with the maximum of 900\,V \cite{arXiv1207}.

\subsection{Experimental setup}
\begin{figure}[tbp]
\centering
\includegraphics[width=0.8\textwidth]{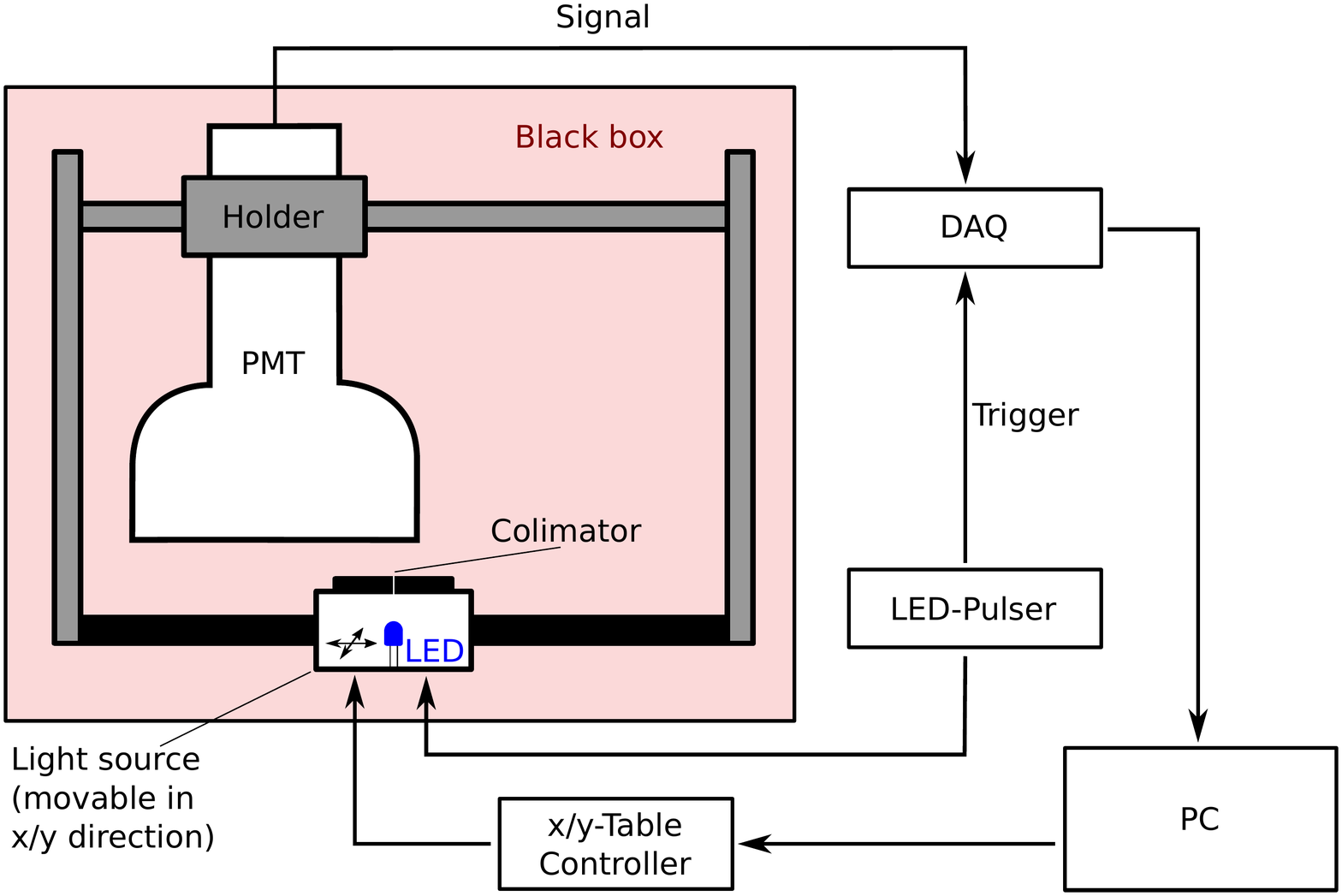}
\caption{\small{Schematic view of the experimental setup,  {\it SandBox}.
The XY-table with the light source and a 0.5\,mm collimator are inserted into the blackbox, and a PMT with its voltage divider is installed into a holder.}}
  \label{setup}
\end{figure}

The measurements, as presented in the following sections, were performed at
room temperature in a light-tight black box which can accommodate four R11410
PMTs fixed in Acrylonitrile Butadiene Styrene holders. Adapters can be
inserted into the holders to fix an R8520 PMT. The custom-made experimental setup ({\it SandBox}) which was used for the measurements,
is shown schematically in Figure~\ref{setup} and described in detail in \cite{BA}. A collimated light source fixed on a two-axis scanner can be controlled from
the outside to move in a plane parallel to the photocathode with a maximal spatial resolution of
15\,$\mu$m. The light source consists of a black anodised light tight aluminium
box containing a blue, 470\,nm wavelength, LED covered with a collimator plate. 
The collimation of the beam is performed by two aligned apertures separated by a
small gap. The apertures used for the presented measurements have 0.5\,mm
diameter. The distance between the collimator and the PMT window is set to 1\,mm.
This results in an enlightened pixel diameter of 0.7\,mm on the photocathode.  
In order to perform the measurements in
a stable and reproducible configuration, they were started not earlier than 60
minutes after switching on the PMTs. The temperature inside the black box
was monitored during the measurements: due to the strong thermal insulation of
{\it SandBox}, achieved by 4\,mm thick tar paper on the inner walls, the observed
temperature fluctuations are lower than 0.5 degrees.

\subsection{Data acquisition and high-luminosity method}

To measure the photo-sensitivity 
uniformity, the emission of photoelectrons is stimulated by
illumination of the photocathode at different positions with collimated 
light from an LED of 470\,nm wavelength.  A pulse generator
(Telemeter TG4001) is used to bias the LED and to simultaneously trigger the
data acquisition. With the high-luminosity (HL) method the light intensity is chosen such that 100\% of the events provide a signal used to
determine the photocathode sensitivity ($\sim$600~photoelectrons). A signal pulse is defined as a charge signal larger than 3 $\times$RMS of the baseline. The light source is fixed on a computer-controlled XY-scanning table~\cite{BA}, that is moved between each measured pixel to scan the entire surface of
the photocathode. The data presented here are acquired using a CAEN
V1724 waveform digitizer with 100 MHz sampling frequency and 40 MHz input
bandwidth. The  employed voltage dividers are identical to those used in XENON100 and XENON1T, and are described in detail by \cite{PhD}. The trigger frequency is 4\,kHz. The waveforms are transferred to a
computer and stored for data processing and subsequent analysis, as well as for visual inspection. A peak processor software scans the digitised waveforms for excursions of 3 times the RMS from the
baseline, integrates the area to obtain the number of electrons contributing to the signal, and histograms the result in a spectrum. The integration window starts when the excursions exceed the defined threshold and ends when the threshold is undercut.
\begin{figure}[btp]
  \centering
  \includegraphics[width=0.6\textwidth]{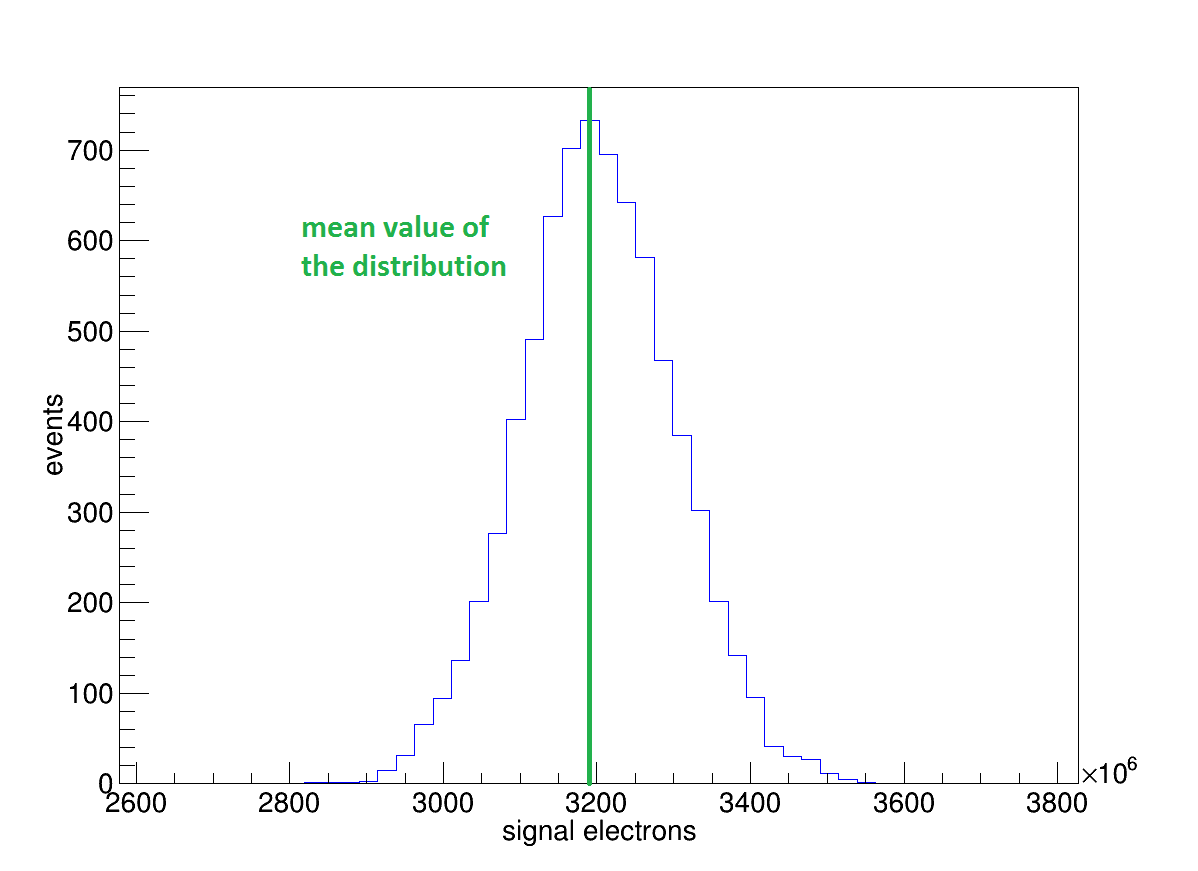}
 \caption{\small{An example of a high-luminosity spectrum of a R11410 PMT. The green vertical line corresponds to the mean value of the the total detected light for one pixel during one measurement. }}
  \label{spe}
\end{figure}
Figure~\ref{spe} shows an example of such an high-luminosity spectrum of a R11410 PMT. The mean of the distribution is proportional to the number of detected photoelectrons, and hence to the sensitivity of the illuminated spot of the photocathode. This
value thus defines the sensitivity of the given spot, with the highest
measured value on the PMT defining 100\% relative sensitivity.


\section{Data analysis}
\label{chap:results}

The result of each analysis can be visualised as a two-dimensional map of relative sensitivities (each normalised by the pixel with the highest sensitivity), a so-called \emph{fingerprint}. 

The non-uniformity is defined by the standard deviation of all pixels in the
fingerprint
\begin{equation}
  s=\sqrt{\frac{1}{n-1}\sum_i^n(x_i-\bar{x})^2},
\end{equation}
where $n$ is the number of pixels, $x_i$ is the 
mean value of the high luminosity spectrum for the pixel $i$, and $\bar{x}$ is
the average value of all pixels in the selected area. 

\subsection{Fingerprints of individual PMTs}

\begin{figure}[btp]
  \centering
  \includegraphics[height=0.41\textwidth]{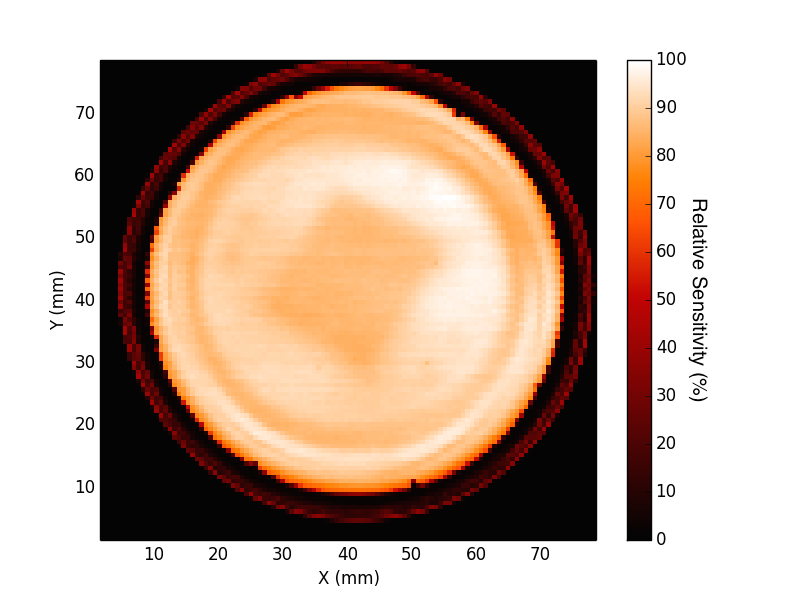}
  \includegraphics[height=0.41\textwidth]{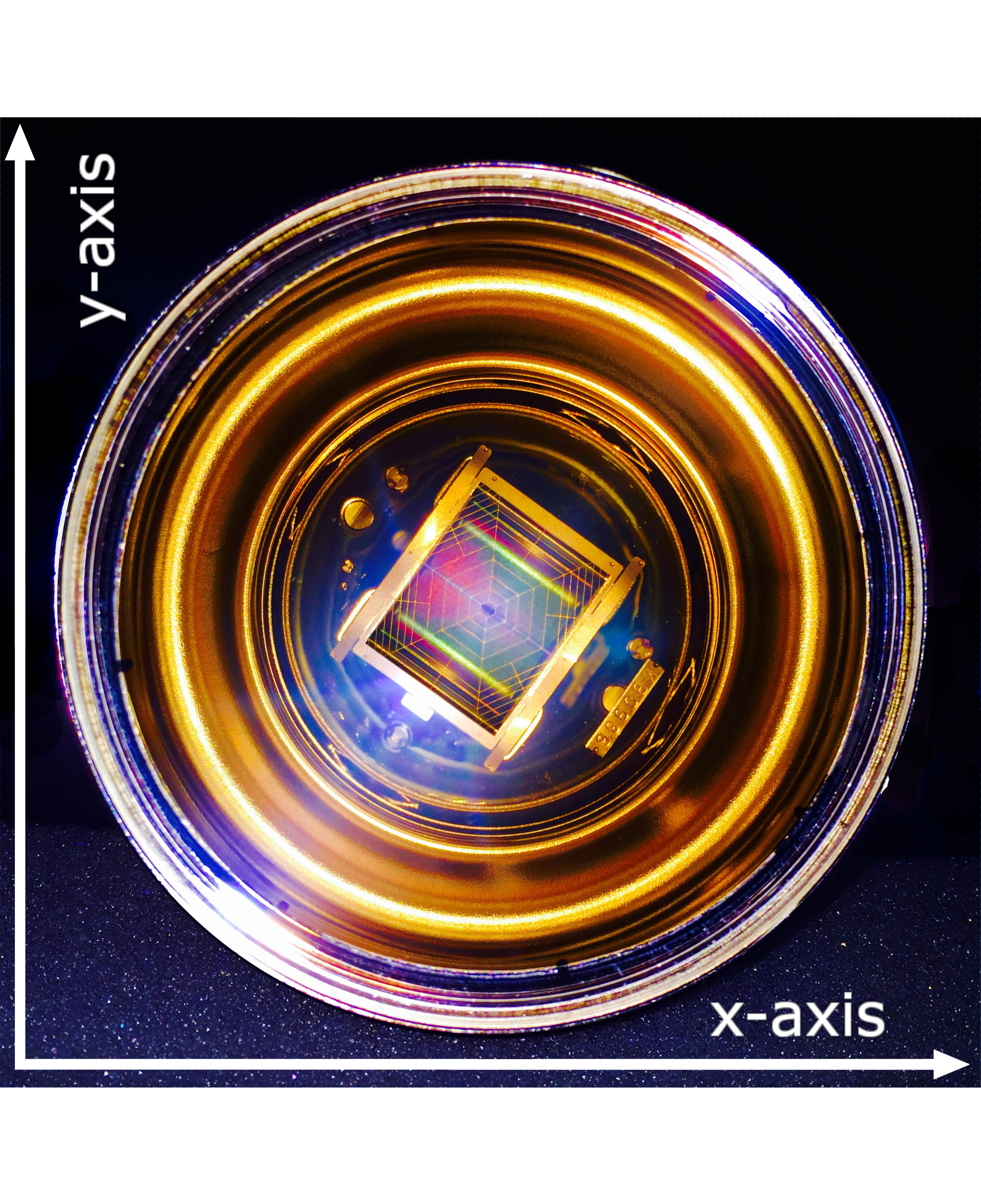}
  \caption{\small{(Left): Planar relative sensitivity of the photocathode of a R11410 PMT normalised by the highest measured value obtained with the high
      luminosity method. A total of 9779 pixels were scanned. (Right): Front view of a R11410 PMT with visible inner structure (focusing dynode, first dynode and screws).}}
  \label{HL_fingerprint}
\end{figure}

\begin{figure}[btp]
  \centering
  \includegraphics[height=0.42\textwidth]{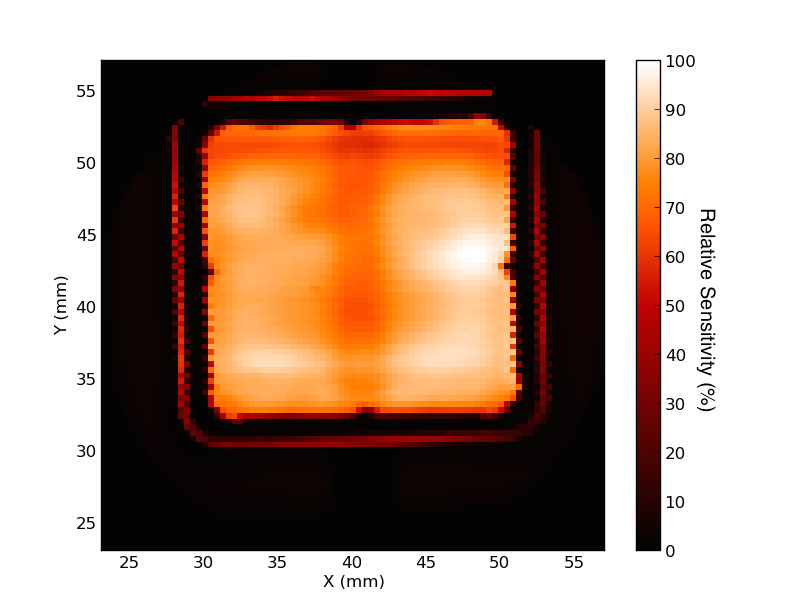}
  \includegraphics[height=0.42\textwidth]{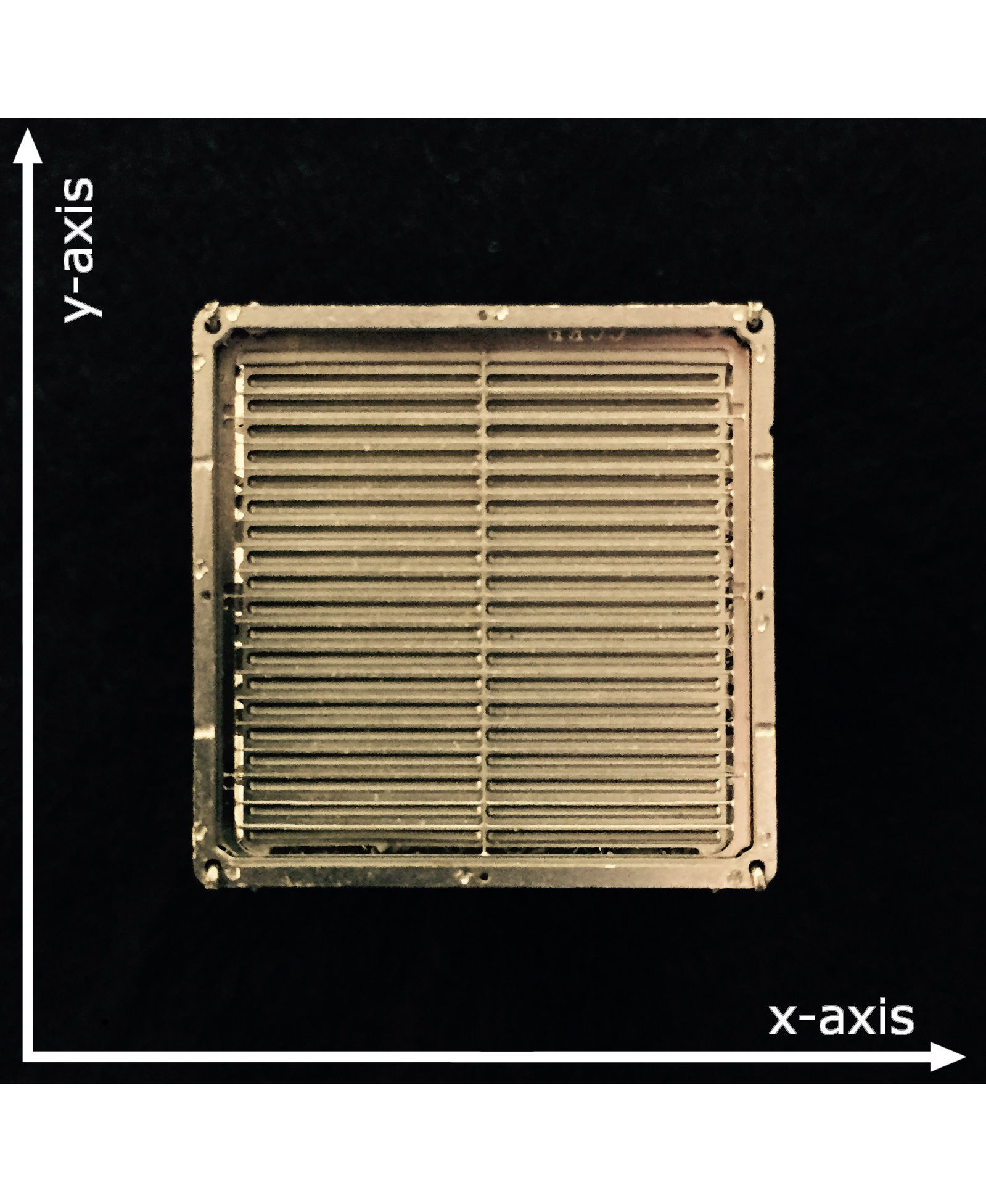}
  \caption{\small{(Left): Planar relative sensitivity of the photocathode of a R8520 PMT
      normalised by the highest measured value obtained with the
      high luminosity method. A total of 4918 pixels were scanned. (Right): Front view of an opened R8520 PMT. The T-shaped holding structure and the first dynode stage are clearly visible.}}
  \label{fig:1inch_fingerprint}
\end{figure}

In Figures~\ref{HL_fingerprint} and \ref{fig:1inch_fingerprint} the fingerprints of a R11410 and a R8520 PMT are shown with a pixel distance of 0.7\,mm, for a total of 9779 and 4918 scanned pixels,
respectively. For each point, $10^4$ events were acquired. The scanning of an R11410 PMT takes 25 hours. The unique properties of each fingerprint are preserved when the PMTs are rotated and rescanned, hence the non-uniformity is not biased by the apparatus.

For the R11410 PMT, the inner structure of the
PMT, shown  in Figure~\ref{HL_fingerprint} (left), becomes visible in the fingerprint.  
This is due to a fraction of incident photons that are not absorbed in the
photocathode and penetrate into the inner volume of the PMT, where they are
reflected on metal surfaces. These photons may hit the photocathode from the rear side
and produce a photoelectron. However, the magnitude of the effect depends on the reflectivity
of these components to the 470\,nm light.  Since the light from the LED is collimated and
hence penetrates the PMT almost perpendicular to the photocathode, the effect
is position-dependent. Pixels with a high reflecting background -- such as
blank flat metal surfaces -- increase the sensitivity of the PMT at that
position, while pixels with lower reflecting background -- such as screw heads
(for example, $(x,y) = (23\,\text{mm}, 45\,\text{mm})$ in
 Figure~\ref{HL_fingerprint}), or the region of the focusing grid and the first
dynode in the center -- show a decreased sensitivity.  From this measurement, it
can be concluded that the dominant variation of the photocathode uniformity is caused by the
reflections of the photons on inner PMT components. 

The inner structure of the R8520 PMT also becomes
visible in the fingerprint. On every dynode, a T-shaped holding structure is
fixed and decreases the sensitivity of the PMT at this position (see Figure~\ref{fig:1inch_fingerprint}). It can be
explained by the fact that some of the incident photons get poorly reflected at the
welding site and some of the emitted electrons from the photocathode are
collected on the welding line instead of the first dynode stage. Due to the absence of a focusing dynode, the photo-electrons can be collected at different positions on the first dynode stage. This results in a position-dependent electron collection efficiency, and hence in regions on the fingerprint that show higher and lower sensitivity.

\subsection{Comparison of different PMTs}

\begin{figure}[btp]
    \centering
    \includegraphics[width=.49\textwidth]{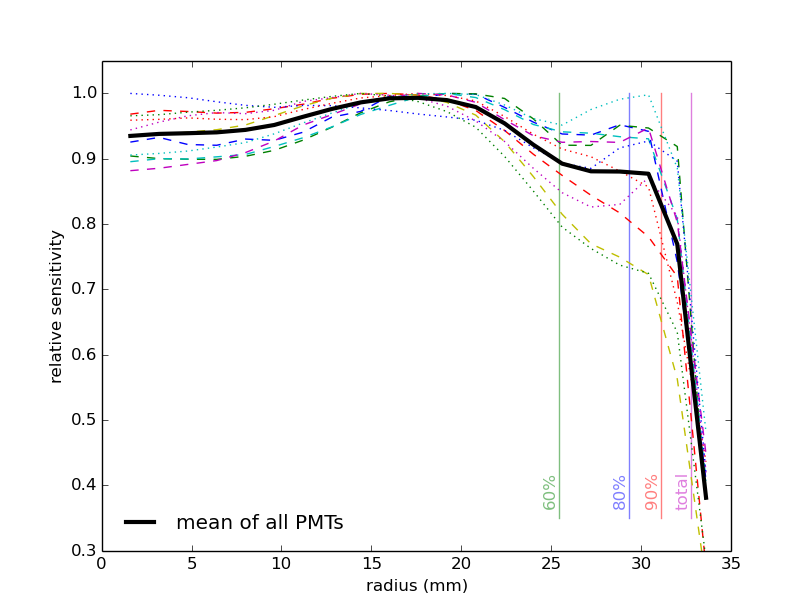}
    \includegraphics[width=.49\textwidth]{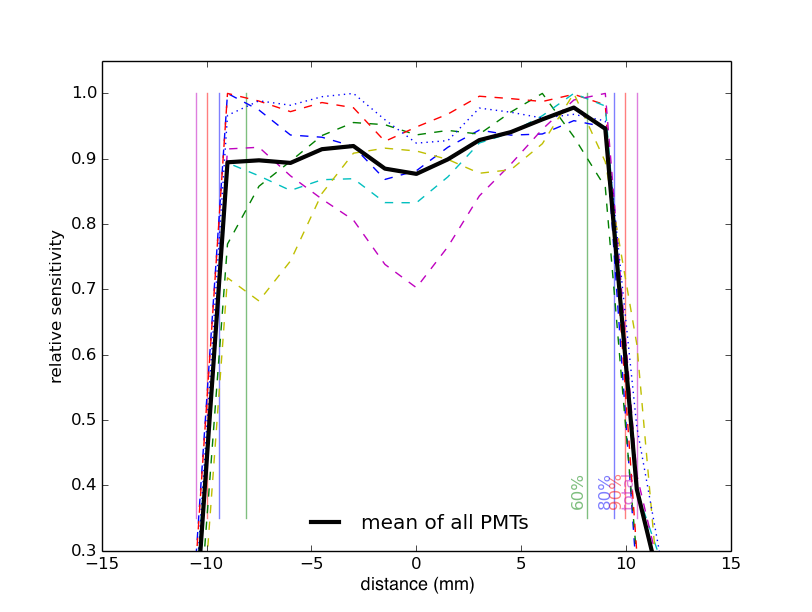}
    \caption{\small{Average relative photo-detection sensitivity (normalised to
      the maximum value of each PMT) of all pixels at a   given radius of eleven R11410 PMTs (left), and average photo-detection
      sensitivity projected along the y-axis of seven R8520 PMTs (right): 
      The solid black curve indicates the average for the scanned PMTs. 
      Thin vertical solid lines mark the radii for
      60\% (green), 80\% (blue), 90\% (red) of the area, as well as for the
      total sensitive PMT area (magenta).}}
    \label{gaudicurves}
\end{figure}

\begin{figure}[btp]
    \centering
    \includegraphics[width=.49\textwidth]{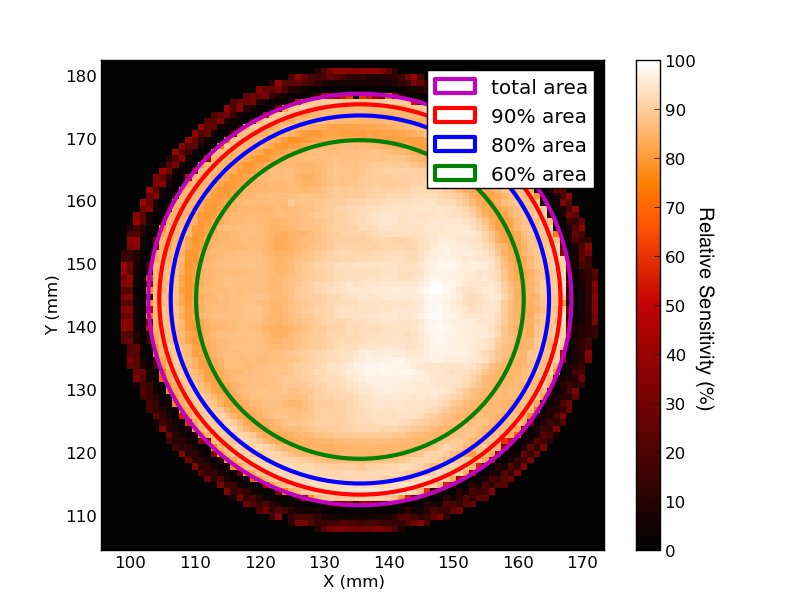}
    \includegraphics[width=.49\textwidth]{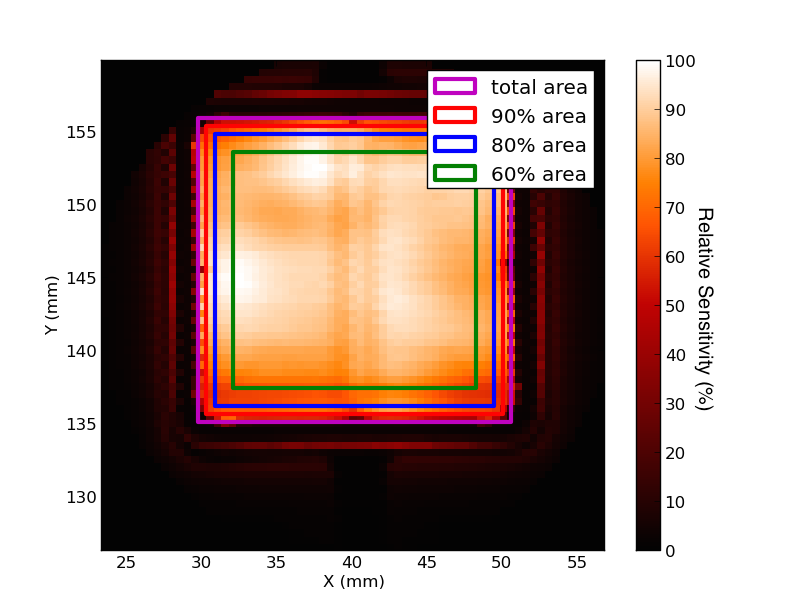} 
    \caption{\small{Different radial or squared selection cuts for 60\%
      (green), 80\% (blue), 90\% (red) of the area, and the total PMT
      area (magenta). Show are a circular R11410 PMT (left) and a square R8520 PMT (right). }}
    \label{fig:selection-cuts}
\end{figure}

Eleven  R11410 PMTs and seven R8520 PMTs were
systematically scanned using the HL method.
Figure~\ref{gaudicurves} shows the radial dependency for the 3-inch PMTs (left) and the dependency projected along the y-axis of the 1-inch PMTs
(right) of the relative sensitivity. For the 3-inch PMTs, the curves show the average of all measured pixels at a particular radius, while for the 1-inch PMTs the average is calculated at a
particular x-position of each PMT. The 3-inch PMTs show a maximum efficiency value
at a radius of about 17\,mm. At the center, the photo-detection efficiency is on
average 7\% lower. For larger radii, there is a larger difference in the trend for different
3-inch PMTs, but the efficiency of all PMTs decreases for a radius larger than $\sim$30\,mm. One side of the 1-inch PMTs is on average 10\% more sensitive than the other. The central region
shows a decreased sensitivity of about 10\% with respect to the most sensitive
border. To quantify the non-uniformity for a given PMT, a certain area on the photocathode is selected.
We use three different selections, as shown in Figure~\ref{fig:selection-cuts}, namely 60\%, 80\% and 90\% of the total area.

\begin{figure}[btp]
    \centering
    \includegraphics[width=0.9\textwidth]{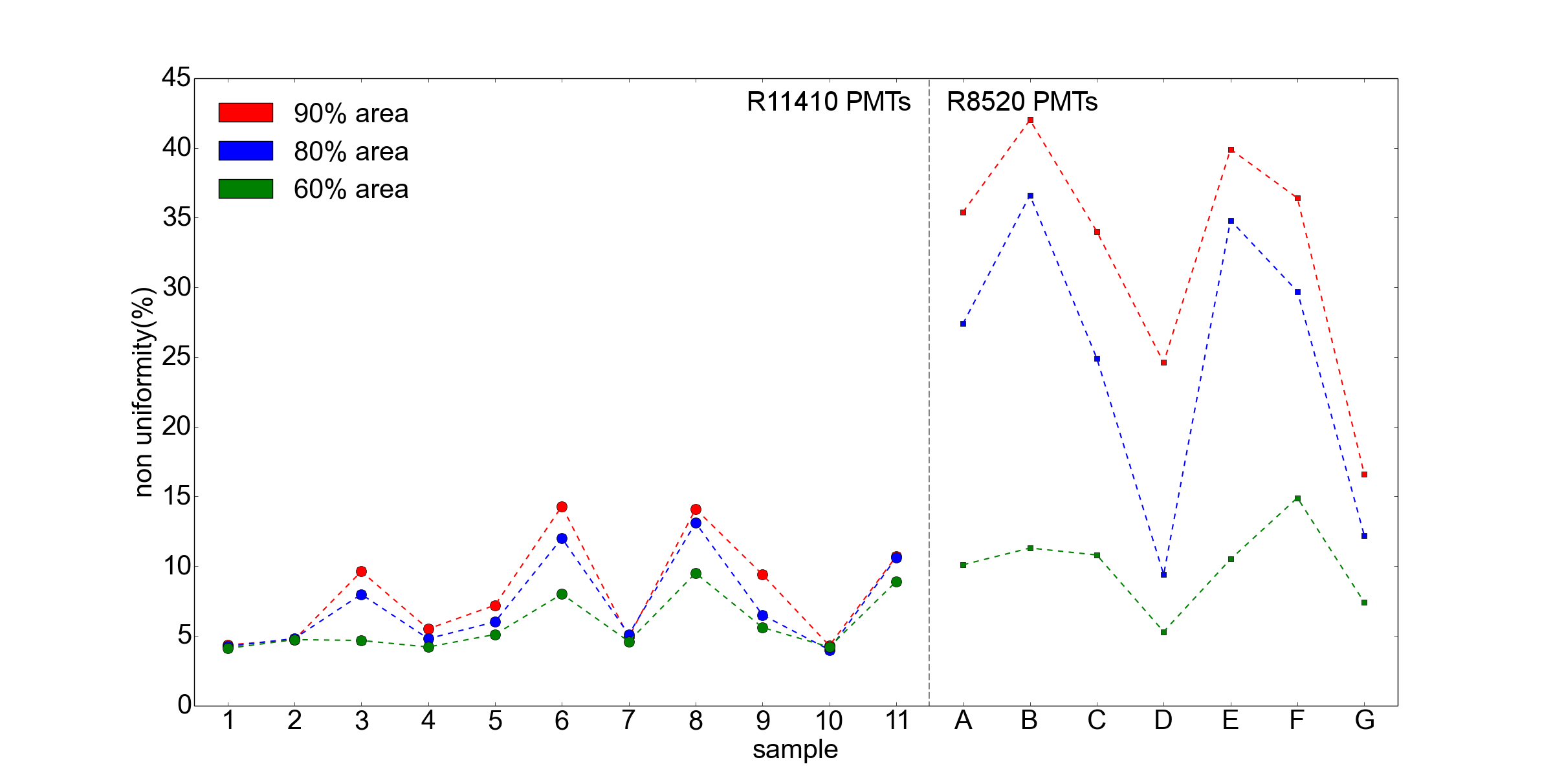}
    \caption{\small{Non-uniformity values   for different active areas on the photocathodes,  
      90\% (red),
      80\% (blue) and
      60\% (green),
      of eleven R11410 PMTs (samples 1-11) and seven R8520 PMTs (samples A-G).} }
    \label{fig:results}
\end{figure}

Figure~\ref{fig:results} shows the non-uniformity values for all scanned PMTs for the three
different area selections. In general, the R11410 sensors show a higher uniformity, even when going to larger
areas, whereas the R8520 PMTs change their uniformity drastically when we take  into account pixels near the sensor edges.

Four of the R11410 PMTs (samples 1, 2, 7, and 10) show an excellent 
uniformity within 60\% of the photocathode area. For these PMTs, the uniformity indeed does
not vary much when  the selected area is increased: for all 3  shown area selections, the measured non-uniformity is less than 5\%. 
The photo-sensitivity non-uniformity of four PMTs (samples 3, 4, 5, and 9) is small for 60\% of the area, namely around 5\%, however it increases for
larger areas, up to 10\%. Three of the PMTs (samples 6, 8, and 11)
have a rather high non-uniformity when selecting 60\% of the area, about 9\%. This value 
increases to 14\% when 90\% of the photocathode area is considered.

Considering 60\% of the photocathode area, the R8520 PMTs show a non-uniformity of about 10\%, which is slightly higher than for the R11410 PMTs. For 90\% of the area, the non-uniformity is between 35\% and 40\%. 
However, two PMTs (samples D and G) show a better non-uniformity of about 6\% and 20\% for 60\% and 90\% of their areas, respectively.


\section{Conclusion}
\label{chap:conclusion}

We measured the relative photo-detection sensitivity at 470\,nm for eleven Hamamatsu R11410 and seven R8520 PMTs, and thus quantified the non-uniformity of their photocathodes. For the 3-inch, R11410 sensors, interesting features were observed: their semi-transparent photocathode is penetrated by a fraction of the photons that may then be reflected on structural parts inside the PMT towards the photocathode,
where they produce a photoelectron. Hence the observed non-uniformity is dominated by the reflectivity of the structural components behind the photocathode.  
We observed  that the non-uniformity is stronger at the edges of the
photocathode surface than at the center region. For the R8520 PMTs, we observed that their response if less uniform, most likely due to the absence of a focusing dynode stage. This effect could be further investigated by simulating the photoelectron paths inside the PMT. Using the non-uniformity value as a quality requirement for the photosensors leads to the conclusion that the 3-inch PMTs show a superior behaviour than the 1-inch sensors. 

In dark matter  detectors using liquid xenon as target, event vertex reconstruction is performed
by computer algorithms which so far assume a perfectly uniform photocathode sensitivity (see for example XENON100 \cite{XE100}).A systematic study of all the PMTs employed in such a detector could potentially improve the uncertainty in event position reconstruction by taking into account the position-dependent sensitivity of each sensor. 
However, since the xenon scintillation light has a wavelength of 175\,nm, the study must be extended to investigate the photo-sensitivity for VUV light. This could be achieved in a similar setup as we described, by using a VUV light source.

\end{document}